\documentclass[journal=jacsat,manuscript=communication,layout=traditional]{achemso}

\usepackage{soul}
\usepackage{graphicx}
\usepackage{epstopdf}

\usepackage[version=3]{mhchem} 
\usepackage{color}
\usepackage{achemso}

\author{Florian Schwarz}
\affiliation[IBM Research - Zurich] {IBM Research - Zurich,
S\"{a}umerstrasse 4, CH-8803 R\"{u}schlikon, Switzerland}
\author{Georg Kastlunger}
\affiliation[University of Vienna] {University of Vienna,
Department of Physical Chemistry, Sensengasse 8/7, A-1090 Vienna,
Austria}
\author{Franziska Lissel} \affiliation[University of
Zurich] {Department of Chemistry, University of Z\"{u}rich,
Winterthurerstrasse 190, CH-8057 Z\"{u}rich, Switzerland}
\author{Heike Riel}
\affiliation[IBM Research - Zurich] {IBM Research - Zurich,
S\"{a}umerstrasse 4, CH-8803 R\"{u}schlikon, Switzerland}
\author{Koushik Venkatesan}
\affiliation[University of Zurich] {Department of Chemistry,
University of Z\"{u}rich, Winterthurerstrasse 190, CH-8057
Z\"{u}rich, Switzerland}
\author{Heinz Berke}
\affiliation[University of Zurich] {Department of Chemistry,
University of Z\"{u}rich, Winterthurerstrasse 190, CH-8057
Z\"{u}rich, Switzerland}
\author{Robert Stadler}
\affiliation[University of Vienna] {University of Vienna,
Department of Physical Chemistry, Sensengasse 8/7, A-1090 Vienna,
Austria}
\author{Emanuel L\"{o}rtscher}
\affiliation[IBM Research - Zurich] {IBM Research - Zurich,
S\"{a}umerstrasse 4, CH-8803 R\"{u}schlikon, Switzerland}
\email{eml@zurich.ibm.com}

\title[\texttt{achemso} Titel]
{High-conductive Organometallic Molecular Wires with Delocalized
Electron Systems Strongly Coupled to Metal Electrodes}

\begin{document}
\noindent \textbf{Abstract} \emph{Besides active, functional
molecular building blocks such as diodes or switches, passive
components as, e.g., molecular wires, are required to realize
molecular-scale electronics.\ Incorporating metal centers in the
molecular backbone enables the molecular energy levels to be tuned
in respect to the Fermi energy of the electrodes.\ Furthermore, by
using more than one metal center and sp-bridging ligands, a
strongly delocalized electron system is formed between these
metallic "dopants", facilitating transport along the molecular
backbone.\ Here, we study the influence of molecule--metal
coupling on charge transport of dinuclear
X(PP)$_2$FeC$_4$Fe(PP)$_2$X molecular wires (PP =
Et$_2$PCH$_2$CH$_2$PEt$_2$); X = CN (\textbf{1}), NCS
(\textbf{2}), NCSe (\textbf{3}), C$_4$SnMe$_3$ (\textbf{4}) and
C$_2$SnMe$_3$ (\textbf{5})) under ultra-high vacuum and variable
temperature conditions.\ In contrast to \textbf{1} which showed
unstable junctions at very low conductance ($8.1\cdot10^{-7}$
G$_0$), \textbf{4} formed a Au-C$_4$FeC$_4$FeC$_4$-Au junction
\textbf{4$'$} after SnMe$_3$ extrusion which revealed a
conductance of $8.9\cdot10^{-3}$ G$_0$, three orders of magnitude
higher than for \textbf{2} ($7.9\cdot10^{-6}$ G$_0$) and two
orders of magnitude higher than for \textbf{3} ($3.8\cdot10^{-4}$
G$_0$).\ Density functional theory (DFT) confirmed the
experimental trend in the conductance for the various anchoring
motifs.\ The strong hybridization of molecular and metal states
found in the C--Au coupling case enables the delocalized
electronic system of the organometallic Fe$_2$ backbone to be
extended over the molecule-metal interfaces to the metal
electrodes to establish high-conductive molecular wires.}\\

\noindent KEYWORDS: Molecular wire, Single-molecule junctions,
electronic transport, break-junctions, organometallic compounds,
density functional theory\\

\noindent Molecular electronics aims at employing single molecules
as functional building blocks in electronic circuits.\ Besides
such active components which provide, e.g., current rectifying or
switching properties, also passive components such as molecular
wires are required for the realization of molecular-scale
electronics.\ Generally, an ideal wire has lowest resistance with
almost linear (ohmic) and length-independent (ballistic) transport
properties.\ For molecular wires, the required high conductance
can in principle be achieved if low injection barriers for
charge-carriers are present at the molecule--metal interfaces, if
molecular orbitals (MOs) are available close to the Fermi energy
of the electrodes, and if a large degree of electronic conjugation
across the backbone is present.\ Already the first task seems to
be difficult to achieve as the most frequently used thiol
anchoring\cite{Bumm1996,Reed1997} suffers from an electronically
weak molecule--metal coupling.\ Additionally, multiple bonding
sites available on the Au surface for the thiol bond give rise to
alternating energy barriers for charge-carrier injection and
result in large fluctuations in the transport properties.\
Therefore other anchoring schemes such as
nitriles\cite{Metzger1997}, isocyanides\cite{Chen1999a},
amines\cite{Quek2007a}, and pyridines\cite{Kamenetska2010} were
investigated.\ Dithiocarbamates\cite{Wrochem2010} were
demonstrated to increase the molecule--metal coupling compared to
previously used single-bond anchors by at least one order of
magnitude, and to simultaneously reduce fluctuations.\ The use of
fullerenes as anchors \cite{martin2008,Fock2011,Lortscher2012a}
seems promising, because of the larger molecule--metal interface
and the affinity of fullerenes for precious
metals.\cite{Joachim1995} However, it turned out that the
transport-limiting barriers shifted from the molecule--metal
interfaces onto the molecular backbone, independently of the
specific connection scheme to the fullerene. \cite{Leary2011} In
contrast to fullerenes with many, but weak sp$^2$ "bonds", the
direct C--Au
 bond showed unprecedented high conductances for oligophenyls
up to 0.9 G$_{0}$,\cite{Chen2011} (for one phenyl ring) close to
the theoretical maximum of 1 G$_{0}$ (with G$_0$ = 2e$^2/h$
$\simeq$ 77 $\mu$S the conductance quantum).\ The C--Au bond can
be established either by extrusion of a trimethyltin
moiety\cite{Chen2011} or post deprotection of a trimethylsilyl
moiety. \cite{Hong2012} Currently, the direct C--electrode bond
seems to be the most promising coupling scheme also for graphene
electrodes\cite{Cao2013,Loertscher2013} if polymerization via the
free termini can be prevented.\

Oligo(phenylene ethynylene)s (OPEs) were considered as one class
of molecular wires as their conjugated backbone enables electron
transport.\ In that respect, C--Au coupled OPEs are currently the
highest conductive molecular wires\cite{Chen2011,Hong2012} with an
exponential conductance decay due to tunneling of approx.\ 1 order
of magnitude per phenyl ring.\ Organometallic
molecules\cite{Ceccon2004} with incorporated metal centers form
delocalized electron systems between two or more metal centers if
appropriate ligand connections over unsaturated C bridges are
chosen \cite{Pevny2010}.\ Furthermore, the MO levels can be tuned
by the metal centers to better align with the Fermi energy of the
leads.\ Motivated by this seminal idea, we have devised dinuclear
Fe complexes\cite{Lissel2014} X(PP)$_2$FeC$_4$Fe(PP)$_2$X
consisting of a [FeC$_4$Fe] backbone with highly delocalized
electronic systems.\cite{Lissel2013} To investigate the effect of
molecule--metal coupling on transport across the [FeC$_4$Fe]
backbone and its influence on the delocalized electronic system,
we varied only the end groups coordinatively or covalently bonded
to the [FeC$_4$Fe] unit.\ All compounds can be considered as
rigid-rod like structures with reduced conformational degrees of
freedom.\ Fig.\ 1 C shows compounds \textbf{1}-\textbf{3} bound
coordinatively via terminal CN, NCS and NCSe end-groups to Au,
whereas the SnMe$_3$ end-capped compounds \textbf{4} and
\textbf{5} (Fig.\ 1 D and 1 E) allow for different covalent
bonding motifs (see SI), e.g.\ to form a direct covalent C--Au
$\sigma$ bond after extrusion of the SnMe$_3$ groups.\ The loss of
the -SnMe$_3$ capping leads to a reduction in length of the
anchoring groups and hence a shorter electrode--electrode distance
for the resulting Au--molecule--Au system.\ The junction's length,
however, determines also the direct electron-tunneling
contribution between the electrodes, a non-negligible electron
path parallel to the molecular-mediated one.\cite{Gotsmann2011}
Accordingly, we couple C$_4$-SnMe$_3$ end groups to the Fe centers
to achieve a length of 2.322 nm (distance between binding Au
atoms) for the Au--\textbf{4$'$}--Au junction that is comparable
to the one of the Au--\textbf{2}--Au (2.257 nm) and
Au--\textbf{3}--Au (2.328 nm) junctions.\ In order to investigate
length-effects on the molecule--electrode coupling, we have
designed additionally compound \textbf{5} with shorter
C$_2$-SnMe$_3$ end groups which forms the Au--\textbf{5$'$}--Au
system with an electrode separation comparable to
Au--\textbf{1}--Au.\ All [FeC$_4$Fe] compounds exhibit a high
charge-delocalization between the two metal centers and can be
oxidized or reduced reversibly in solution with up to three
oxidation states at relatively low potentials ($<$ 1.0
V).\cite{Lissel2013,Lissel2014} (see SI).\

To perform transport measurements, we use electron-beam-structured
break-junctions (Fig.\ 1 B) that are mechanically actuated in a
three-point bending mechanism (Fig.\ 1 A) operated in an
ultra-high vacuum environment (UHV; pressure $p<$ 2 $\cdot$
10$^{-9}$ mbar) and at variable temperature (10 K $< T <$ 300
K)\cite{Loertscher2007} (see SI for details).\ Statistical data
acquisition is performed by taking several hundred $I$--$V$
characteristics curves in subsequent junction forming and breaking
cycles.\cite{Loertscher2007} Due to microscopic surface
reconfigurations under the applied high fields and at elevated
temperatures, only the opening data is considered.\ We first
report on the transport properties of the compounds \textbf{1} -
\textbf{5} taken at room-temperature (300 K).\ The measurement of
compound \textbf{1} upon initial junction closing and subsequent
opening and closing cycles under a fixed bias of 50 mV resulted in
histograms that showed less distinct molecular signatures with a
small conductance accumulation located at around $8.1\cdot10^{-7}$
$G_{0}$ (see SI).\  $I$--$V$ data acquisition was not possible due
to highly unstable junctions.\ In contrast, compounds \textbf{2},
\textbf{3}, \textbf{4} and \textbf{5} (transformed into
\textbf{4$'$}, and \textbf{5$'$} respectively, upon attachment to
the Au electrodes) gave reproducible $I$--$V$ data upon repeated
opening of the junction.\ The $I$--$V$ data gathered was then
mathematically derived to obtain (differential) conductance vs.\
voltage, $G_{Diff}$-$V$, curves.\ The entity of all these opening
curves is displayed as a "density plot" in the left column of
Fig.\ 2 with the color code representing the grade of
accumulation.\ The data contains 1033 $I$--$V$ characteristics
taken for \textbf{2} (with a junction forming probability of
70$\%$), 812 for \textbf{3} (70$\%$), 636 \textbf{4} (98$\%$), and
1929 for \textbf{5} (70$\%$) as acquired during the identical
measurement protocols of comparable cycle numbers.\ Based on the
most probable accumulations, we have selected individual
$G_{Diff}$-$V$ characteristics (transparent blue curves) to
display the functional behavior of individual curves.\ In
addition, conductance histograms were constructed by taking the
conductance data at $\pm$ 1.0 V from the opening curves (see SI
for histograms extracted at other voltages and in absence of
molecules).\ According to our measurement approach, the electrodes
are brought in very close contact (approx.\ 0.1 nm) during every
cycle, which results either in the formation of a direct Au--Au
contact or multi-molecular junctions, depending primarily on the
diffusion of surface Au atoms under the applied high field.\
Hence, the close-contact or high-conductance regime of (0.08 -
5.0) G$_0$ is therefore considered as not appropriately controlled
at room temperature and henceforth indicated by a blue shaded
background in the right column of Fig.\ 2.\

Fig.\ 2A shows one broad and two narrow accumulations of
$G_{Diff}$--$V$ data for \textbf{2}.\ The corresponding
conductance peaks in the histogram are located at 0.95 G$_{0}$,
$1.5\cdot10^{-1}$ G$_{0}$ and $7.9\cdot10^{-6}$ G$_{0}$ as
displayed in Fig.\ 2B.\ The first distribution represents Au--Au
QPCs that are formed repeatedly during the measurement process.\
The most dominant and hence most probable distribution at
$7.9\cdot10^{-6}$ G$_{0}$ is attributed to the formation of a
Au--\textbf{2}--Au junction.\ In contrast, transport measurements
of compound \textbf{3} reveal no clear accumulation in the
$G$--$V$ data (Fig.\ 2 C).\ Instead, a spread in the
$G_{Diff}$--$V$ data from $10^{-5}$ G$_{0}$ to $10^{-2}$ G$_{0}$
is found.\ The conductance histogram confirms this finding by a
broad peak located at $3.8\cdot10^{-4}$ G$_{0}$.\ Much more
distinct are the results for compound \textbf{4}, where three
peaks are found at $0.86$ G$_{0}$, $8.9\cdot10^{-3}$ G$_{0}$ and
$9.6\cdot10^{-7}$ G$_{0}$ (Fig.\ 2 F), as could also be presumed
from the $G$--$V$ distribution (Fig.\ 2 E).\ Here, the first peak
again originates from Au--Au metal junctions, whereas the second
and third one are due to the formation of a Au--\textbf{4$'$}--Au
junction.\ From the peak height, i.e.\ the relative occurrence, we
preliminarily conclude that the most probable conductance is
$7.9\cdot 10^{-6}$ G$_{0}$ for \textbf{2}, $3.8\cdot 10^{-4}$
G$_{0}$ for \textbf{3}, and $8.9\cdot 10^{-3}$ G$_{0}$ for
\textbf{4$'$} (all taken at 1 V).\ Besides the difference in the
conductance maxima, also the spread in conductance differs clearly
for the three different anchor groups being studied.\ For NCS and
NCSe anchoring, the widths of the conductance histograms are
approx.\ 3-4 orders of magnitude (e.g., G$_{3,high}$/G$_{3,low}$ =
4 $\cdot 10^{3}$, estimated from the Full Width at Half Max (FWHM)
of a Gaussian-like peak), and much less for direct C--Au
anchoring, approx. 1-2 orders of magnitude
(G$_{4,high}$/G$_{4,low}$ = 2.5$\cdot 10$).\ This smaller
conductance variation is also found for the second C--Au coupled
and shorter Au-- \textbf{5$'$}--Au system as displayed in Fig.\ 2
G and H, which show an even higher conductance of $1.3\cdot
10^{-2}$ G$_{0}$.\

At room temperature, the MOs energy level are usually broadened
and the Fermi energy of Au is broadened too, leading to rather
monotonic and continuous $I$--$V$ characteristics as displayed in
Fig.\ 2 for all compounds.\ In contrast, the MOs usually become
apparent in $G$--$V$ characteristics at low temperatures,
typically at less than 100 K, because of the reduced thermal
broadening.\ We therefore investigated the transport properties
exemplarily for \textbf{2} and \textbf{4$'$} at low temperatures
(Fig.\ 3).\ The data exhibits a symmetric conductance gap of
approx.\ 0.8 V for \textbf{2}, independent on the temperature (the
data contains 120 $I$--$V$ characteristics, 40 taken at 30 K, 50 K
and 100 K each).\ In the low-voltage range up to $\pm$ 0.25 V, no
MOs are available for electrons to tunnel through.\ At higher
bias, however, the current starts to increase as frontier MOs
(according to DFT the HOMO, see below) get into resonance.\ As can
be seen best in the $G$-$V$ representation, where the resonant MOs
are represented by peaks, they are located at -0.85 V, -0.39 V,
0.39 V, and 0.87 V.\ They are spaced symmetrically with respect to
bias polarity, as it is expected for symmetric molecules and
symmetric coupling.\ In addition to the conductance gap and the
appearance of discrete MO resonances in \textbf{2}, many $I$--$V$
characteristics with the appearance of hysteretic conductance
switching are found (see SI).\ All these findings differ strongly
to those for compound \textbf{4$'$}, where only monotonous curves
without a conductance gap were recorded at low temperatures.\
Fig.\ 3 shows 100  $I$--$V$ (C), and $G_{Diff}$-$V$ (D)
characteristics of \textbf{4$'$}, taken at 50 K (similar data for
30 K and 100 K):\ Besides the absence of discrete MO peaks, the
transport properties are more linear and the current levels are 3
to 4 orders of magnitude higher.\

To study the MO alignment and landscape, we performed Density
Functional Theory (DFT) calculations with a PBE XC-functional
within a NEGF-DFT framework\cite{Brandbyge2002,Xue2002,Rocha2005}
using the GPAWcode\cite{Mortensen2005,Enkovaara2010} to compute
transmission probabilities, $T(E)$.\ In order to account for
selfinteraction errors and image charge effects present in DFT
with local XC-functionals we applied a scissor operator (SO),
according to Quek et al. \cite{Quek2007a}, to the weaker coupled
molecules \textbf{1} to \textbf{3} (see SI).\ All DFT calculations
were carried out without treating spin polarisation as a degree of
freedom since previous tests on Fe complexes with the same ligand
field revealed the low spin configuration to be the ground state.\
The results of the DFT calculations for the transmission functions
T(E) and eigenenergies of the respective orbitals HOMO and HOMO-1
relative to E$_F$ are presented in Fig.\ 4A and B for the
compounds \textbf{1} to \textbf{5$'$}.\ Fig.\ 4C displays
calculated $I$--$V$ curves that were obtained from the
transmission functions T(E) in a rigid band approximation where
the bias dependence of T(E) is disregarded, as $I = \frac{2e}{h}
\int_{-\infty}^{+\infty}T(E)[f_1(E)-f_2(E)]dE$ with $f_{1,2}$
     as the respective Fermi functions for the two electrodes at
     50 K and their chemical potentials shifted by $\pm eV/2$.\
The figure illustrates the relation between the energetic position
of those two MOs and the characteristic double peaks in the
transmission.\ Furthermore it shows the spatial distribution of
these two MOs.\ Both the eigenvalues and the shape of the relevant
MOs are similar for all systems, consisting of $\pi$-orbitals
delocalized over the entire molecular backbone and containing
equal amounts of both Fe d-states.\ For each system, the HOMO and
HOMO-1 differ only in the sense that they are rotated by
90$^{\circ}$ to each other, which might indicate an energetic
degeneracy of the two states. However the rotational symmetry is
slightly disturbed by the (PP)$_2$ ligands on the Fe centers
explaining the small energetic splitting and therefore the
appearance of a double-peak structure in the transmission
function.\ The conductance at zero bias, which is given in Fig.\ 5
B) and compared to experimental findings, is mainly influenced by
the tails of the HOMO and HOMO-1 peaks, leading to quite different
values among the compounds investigated.\ Although the
metal--molecule coupling is quite high for all anchor groups, the
two C--Au end groups surpass the others with rather strong
covalent bonding, which leads not only to broad peaks in the
transmission function, but also to a more distinct energy shift of
the peaks towards $E_{F}$ caused by hybridization of the MOs and
the lead bands.\ It can be seen that the aligned MO eigenenergies
for the different anchor schemes are rather similar with the
exception of compound 1, thereby ruling out structural variations
in the charge transfer\cite{Stadler2006, Stadler2006a,Stadler2010}
as a possible source for the differences in the transmission peak
energies for compound 2-5', and leaving only variations in the
hybridization strengths as explanation.\ As a consequence, even
the rather long C$_4$ anchors of \textbf{4$'$} lead to a higher
conductance than the coordinatively bonding end groups CN, NCS and
NCSe, although the rate of coherent tunneling decreases rapidly
with the Au--Au distance in a molecular junction.\ Similar to the
arguments for the superior conductance provided by the C--metal
end groups, also the conductance ordering for the thiol and
selenium anchors can be rationalized by the fact that the
electronic coupling strength of Se--Au exceeds that of
S--Au\cite{Patrone2002,Yaliraki1999a} due to a larger overlap of
the wavefunctions.\
 We start the discussion of experimental and
theoretical findings with compound \textbf{1}.\ The presence of
only weak and rather unlikely molecular signatures (of
$8.1\cdot10^{-7}$ $G_{0}$ at 50 mV bias) in the low-bias transport
data of compound \textbf{1}, can have several reasons: First, the
conductance of compound \textbf{1} is either below our
experimental resolution ($\ll$ $1.0\cdot10^{-8}$ $G_{0}$), or,
second, the CN binding to Au is weak and the resulting
Au-\textbf{1}-Au junction is not stable under high bias, or,
third, the bulky ligands prevent the terminals to bind to the Au
electrodes due to the short distance to the Fe center.\ For
compounds \textbf{2}, \textbf{3}, \textbf{4$'$} and \textbf{5$'$},
the room-temperature experiments worked reproducibly and the
conductance data displayed in Fig.\ 2 shows values that range from
slightly larger than 1 G$_0$ down to 10$^{-8}$ G$_0$.\ It is hence
ensured that all possible configurations during the junction
forming and breaking procedure, from fully open Au contacts to
Au--molecule--Au junctions and direct Au--Au QPCs were probed.\
The QPC peak at 1 G$_{0}$ confirms that the electrodes completely
touched (at least in some of the cycles) in the required gentle
way, i.e., not fusing the contact entirely.\ The data gathered,
noticeably, represents conductances of all possible electrode
distances.\ In case of \textbf{2}, a broad peak with a maximum at
$7.9\cdot10^{-6}$ G$_{0}$ is formed.\ The fluctuations giving rise
to this broad peak are typically generated by variations in the
S--Au bond as multiple bonding sites (top, hollow, bridge etc.)
are available on the Au surface.\ An even wider peak is found for
the Se--Au bond of compound \textbf{3}, indicating multiple
bonding sites with fast binding kinetics and low transition states
for site exchange that do not necessarily need thermodynamic
activation for the weaker Se--Au (binding energy of 0.516 eV
compared to 0.669 eV for S--Au) bond.\ For both C--Au coupled
compounds \textbf{4$'$} and \textbf{5$'$}, much narrower
conductance accumulations are found.\ In the DFT calculations, the
top position was identified to be the energetically most stable
configuration.\ As a consequence, the C--Au anchors of compounds
\textbf{4$'$} and \textbf{5$'$} are supposed to be in their
equilibrium bonding-site configuration even under mechanical
manipulation of the junction, which results in narrow conductance
histogram peaks.\ In the transport data of compound \textbf{4$'$}
(and weaker also in case of \textbf{5$'$}), a second, broader but
smaller peak compared to the main peak at $8.9\cdot10^{-3}$
G$_{0}$ is found at $9.6\cdot10^{-7}$ G$_{0}$.\ The appearance of
a second peak at a lower average conductance for \textbf{4$'$}
(and similar also for \textbf{5$'$}) is presumed to originate from
the various bonding scenarios of the C end group: Incomplete
cleavage of the SnMe$_3$ capping, formation of chemically
reasonable alkynyl vinylidene trimethyltin species
[(-C$\equiv$C)(SnMe$_3$)C=C)] upon binding to the gold electrode
resulting in the formation of a carbene type bond to the Au
electrode ([Au-C$_4$FeC$_4$Fe-C$\equiv$C(SnMe$_3$)C=C=Au] =
Au--\textbf{4$"$}--Au) (see SI), transport through one of the
bis(diethylphosphino)ethane ligands (as one or two arms of the
phosphine ligands could lift-off to form Fe-PCH$_2$-CH$_2$-P
$\rightarrow$ Au) and non-cleaved end groups cappings.\
Alternatively in our understanding, also reductive C--C coupling
forming a dimerized Au--C$_4$FeC$_4$FeC$_8$FeC$_4$FeC$_4$--Au
(Au--\textbf{4$'$}-\textbf{4$'$}--Au) junctions (similarly for
\textbf{5$'$}) can occur.\ As such details of the junction
configuration are experimentally not directly accessible, the
conductances of the Au-\textbf{4$'$}-\textbf{4$'$}-Au and
Au-\textbf{5$'$}-\textbf{5$'$}-Au dimer junctions and the
vinylidene-coupling case were exemplarily calculated (see SI).\ A
conductance of $1.05\cdot10^{-5}$ G$_{0}$ was found for the dimer
junction Au--\textbf{4$'$}-\textbf{4$'$}--Au.\ In the transmission
function of the dimer, the slope at the Fermi level is relatively
high, which means that a small energy shift of 0.1 eV would result
in a lower calculated conductance.\ This notion is in agreement
with the experimental finding as such a small shift in energy
could also be argued to arise from deficiencies of DFT such as gap
underestimation.\ Due to the good agreement between DFT and
experiments for both the 'monomer' and the 'dimer' compounds, we
conclude that spontaneous dimerization is most likely the origin
for the low-conductance peaks of compounds \textbf{4} (and also
\textbf{5}), in agreement with the observation of dimerization in
SnMe$_3$-capped oligophenyls with C--Au anchors.\cite{Chen2011} A
dimerization explains further why the contacting traces for
molecular signatures are 5-7 times longer for the low-conductance
I-Vs compared to the high-conductance $I$-$V$s (see SI).\

When comparing the main peaks in the conductance data at high bias
(1.0 V) or low bias (0.2 V, see SI) of \textbf{2}, \textbf{3},
\textbf{4$'$} and \textbf{5$'$} measured at 300 K, a good
qualitative agreement with DFT at zero-bias is found as directly
compared in Fig.\ 5 B).\ The zero-bias conductance according to
DFT and the low-bias current in the experiments are both much
higher for \textbf{4$'$} or \textbf{5$'$} than for \textbf{2} and
\textbf{3}, which indicates that the LDOS is much higher for the
C--Au coupled systems than that of the others.\ The orbital
distribution indicates that a strong hybridization of MOs and
metal states takes places at the molecule--metal interfaces in the
C--Au coupled system as evidenced by the difference in the HOMO's
amplitude on the bonding site as obtained from DFT data
highlighted by circles in Fig.\ 5 D.\ This hybridization shifts
HOMO and HOMO-1 closer to $E_F$, leading to an earlier onset in
electron transport as evidenced by the low-temperature transport
properties where the conductance gap has even vanished (Fig.\ 3).\
Injection barriers estimated from minima in the
transition-voltage-spectroscopy representation
(ln(I/V$^2$)--(1/V); see SI) reveal a similar barrier height of
(1.75 $\pm$ 0.3) 1/V for \textbf{4$'$} and (1.85 $\pm$ 0.3) 1/V
for \textbf{5$'$} in contrast to (4.2 $\pm$ 1.5) 1/V for
\textbf{3}, and (5.5 $\pm$ 1.5) 1/V for \textbf{2} at 300 K.\ The
strong hybridization of metal and molecular states established by
the C--Au coupling might further be the reason why the hysteretic
switching behavior found at low temperatures for the weakly
coupled compound \textbf{2}
 (see SI) was not revealed in the strong C--Au
coupled compound \textbf{4$'$} as the MOs are more pinned and
intrinsic functionality might be prohibited.\ The energetic
positions of the frontier MOs found for compound \textbf{2} at
around $\pm$ 0.4 V at low temperatures are in quantitative
agreement with the energetic difference between HOMO and $E_F$
calculated by DFT to be around 0.25 eV - 0.30 eV as illustrated in
Fig. 4B.\ These values are around 100 meV smaller than the MO
energies in Fig.\ 3B which is due to the mean field character of
DFT with semi-local exchange correlation functionals which do not
capture many body effects.\cite{Geskin2009,Stadler2012}

Compared with trimethylsilyl-\cite{Hong2012} or
trimethyltin-capped oligophenyls with a direct Au--benzene
attachment,\cite{Chen2011a} the conductance of compound
\textbf{4$'$} is more than ten-fold higher if similar wire
lengths, $l$, (approx.\ 2 nm) are taken into account.\ When
comparing with organometallic ruthenium(II)
bis($\sigma$-arylacetylide) complexes with SCN-Au coupling
\cite{Kim2007,Luo2011}, the conductance of \textbf{4$'$} is more
than one order of magnitude higher.\ For trimethyltin-capped
polyphenyls with  additional carbon atoms in the Au--C--benzene
bonds,\cite{Chen2011} a conductance of $1.4\cdot10^{-2}$ G$_{0}$
was found for 4 phenyl units, similarly high as the one of
compound \textbf{4$'$}.\ When taking the dimer system
Au--\textbf{4$'$}--\textbf{4$'$}--Au into account, we can create a
preliminary length-dependence for the conductance decrease with
wire length ($G$ $\propto$ e$^{-\beta/l}$) of the Fe-based
organometallic wires to compare with state-of-the-art molecular
wires (see SI).\ The decay constants of $\beta$ = 4.4 nm$^{-1}$
(determined by experimental values at 200 mV or 1.0 V) and $\beta$
= 3.5 nm$^{-1}$ (DFT at zero bias) are both higher than for the
organometallic ruthenium(II) bis($\sigma$-arylacetylide)
complexes\cite{Kim2007,Luo2011} ($\beta$ = 1.02 - 1.64 nm$^{-1}$)
or purely organic oligothiophenes\cite{Yamada2008} ($\beta$ = 1.0
nm$^{-1}$) with lowest decay constants reported so far.\ The
values estimated and calculated are closer to decay constants for
phenyls coupled via C--Au\cite{Chen2011} ($\beta$ = 4.0 - 6.0
nm$^{-1}$).\ A full experimental study of oligomeric
organometallic molecules with 1 to 4 repeating Fe units, however,
has to confirm this preliminary estimation.\

In summary, we have theoretically and experimentally investigated
the influence of molecule--metal coupling on the electron
transport properties of dinuclear Fe complexes.\ We varied the
molecule--metal coupling systematically by using different
anchoring schemes, such as CN, NCS, NCSe, C$_2$SnMe$_3$ and
C$_4$SnMe$_3$ with the latter two end groups leading to a direct
C--Au bond after SnMe$_3$ extrusion.\ Whereas the CN termination
did not result in stable junctions, all other end groups yielded
reproducible transport junctions that enabled the determination of
the room-temperature coupling strengths, which follow the order
$\Gamma_{NCS-Au}$ $<$ $\Gamma_{NCSe-Au}$ $<$ $\Gamma_{C_4-Au}$ $<$
$\Gamma_{C_2-Au}$, in qualitative agreement with DFT
calculations.\ Moreover, the reproducible binding of the C-Au
motif upon extrusion or migration of the SnMe$_3$ end-group was
demonstrated to occur also at low temperatures (50 K), leading to
the formation of high-conductive molecular wires.\ Overall, the
class of organometallic compounds with delocalized electron
systems between two and more metal centers is a promising concept
to achieve long and highly conductive wires due to an extension of
the electronic system of the [FeC$_4$Fe] unit over the
molecule--metal interfaces to the electrodes by strong
hybridization.\ Beyond that, organometallic compounds are an
attractive framework for the integration of intrinsic
functionality for future applications such as redox activity for
conductance switching and memory
application.\\

\noindent \textbf{ASSOCIATED CONTENT} \\

\noindent Supporting Information on the synthesis of the
compounds, the experimental setup, control measurements,
histograms at other voltages and additional DFT calculations.\
This material is available free of charge via the
Internet at http://pubs.acs.org.\\

\noindent \textbf{AUTHOR INFORMATION}\\

\noindent Corresponding Author for chemistry: hberke@chem.uzh.ch
(H.B.), venkatesan.koushik@chem.uzh.ch (K.V.); DFT calculations:
robert.stadler@univie.ac.at (R.S.);
experiments: eml@zurich.ibm.com (E.L.)\\

\noindent \textbf{NOTES}\\

\noindent The authors declare no competing financial interest.\\

\noindent \textbf{ACKNOWLEDGMENTS}\\

\noindent We are grateful to B.\ Gotsmann, V.\ Schmidt, and W.\
Riess for scientific discussions, and to M.\ Tschudy, U.\
Drechsler and Ch.\ Rettner for technical assistance.\ Funding from
the National Research Programme "Smart Materials" (NRP 62, grant
406240-126142) of the Swiss National Science Foundation (SNSF) and
the University of Z\"{u}rich is gratefully acknowledged. G.K. and
R.S. are currently supported by the Austrian Science Fund FWF,
project Nr. P22548 and
 are deeply indebted to the Vienna Scientific Cluster VSC, on whose
 computing facilities all DFT calculations
 have been performed (project Nr. 70174).\\

\bibliographystyle{apsrev}
\bibliography{Dinuclear_Fe_revised_V2}

\begin{figure*}[htbp]
    \begin{center}
   \resizebox{1.5\columnwidth}{!}{\includegraphics{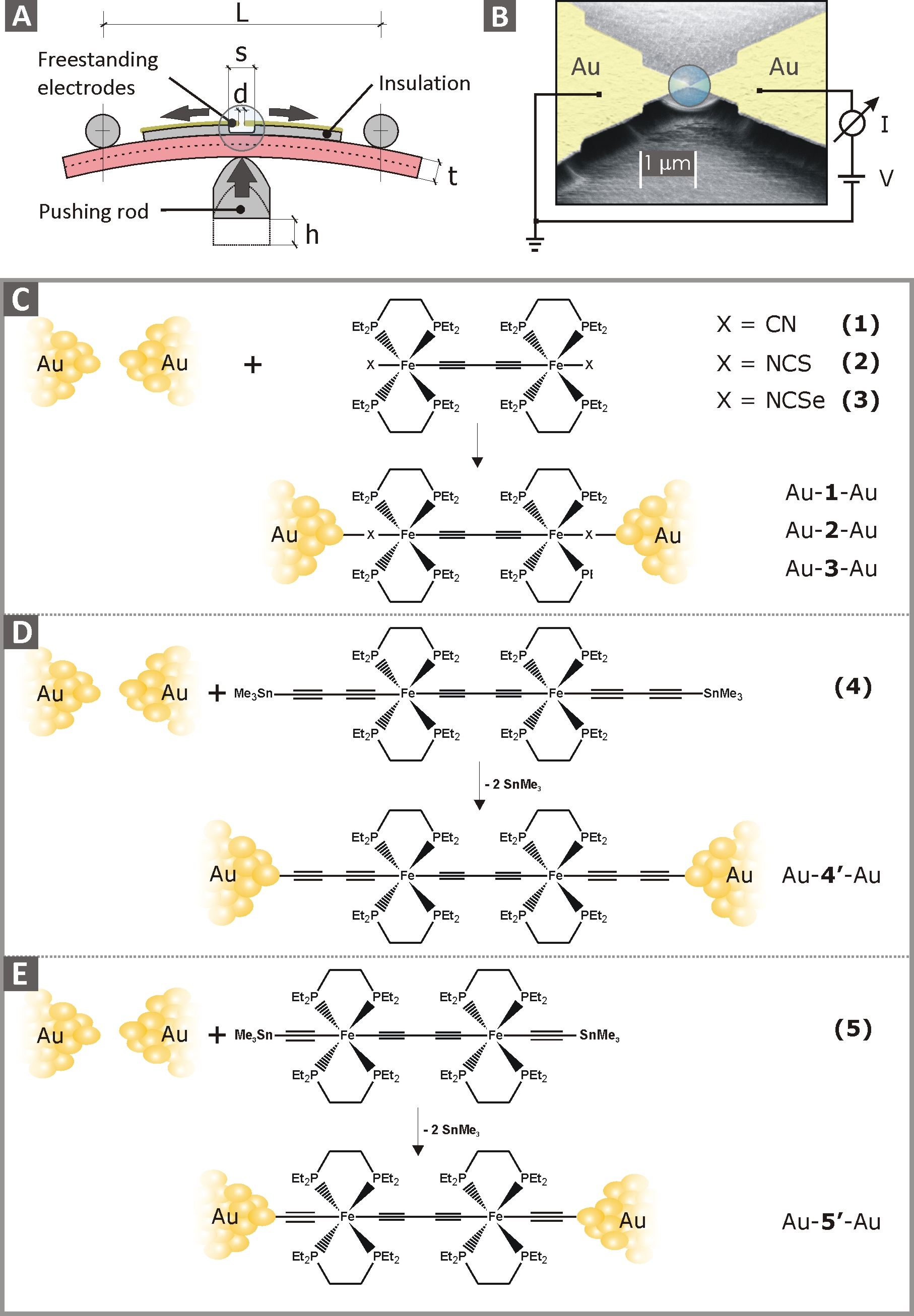}}
    \caption{\label{fig:1} A) Operation principle of a mechanically
    controllable break-junction.\ B)
    Scanning electron microscope (SEM) image of a micro-structured sample.\ C)
Compounds
    \textbf{1} - \textbf{3} with corresponding reaction schemes upon coupling
    to Au electrodes.\ In contrast to compounds \textbf{1} - \textbf{3}, the
SnMe$_3$ end groups
    of \textbf{4} and \textbf{5} cleave off and direct C--Au bonds are formed
    yielding the Au--\textbf{4$'$}--Au (D) and the Au--\textbf{5$'$}--Au
junction (E), respectively.}
\end{center}
\end{figure*}

\begin{figure*}[htbp]
    \begin{center}
    \resizebox{1.6\columnwidth}{!}{\includegraphics{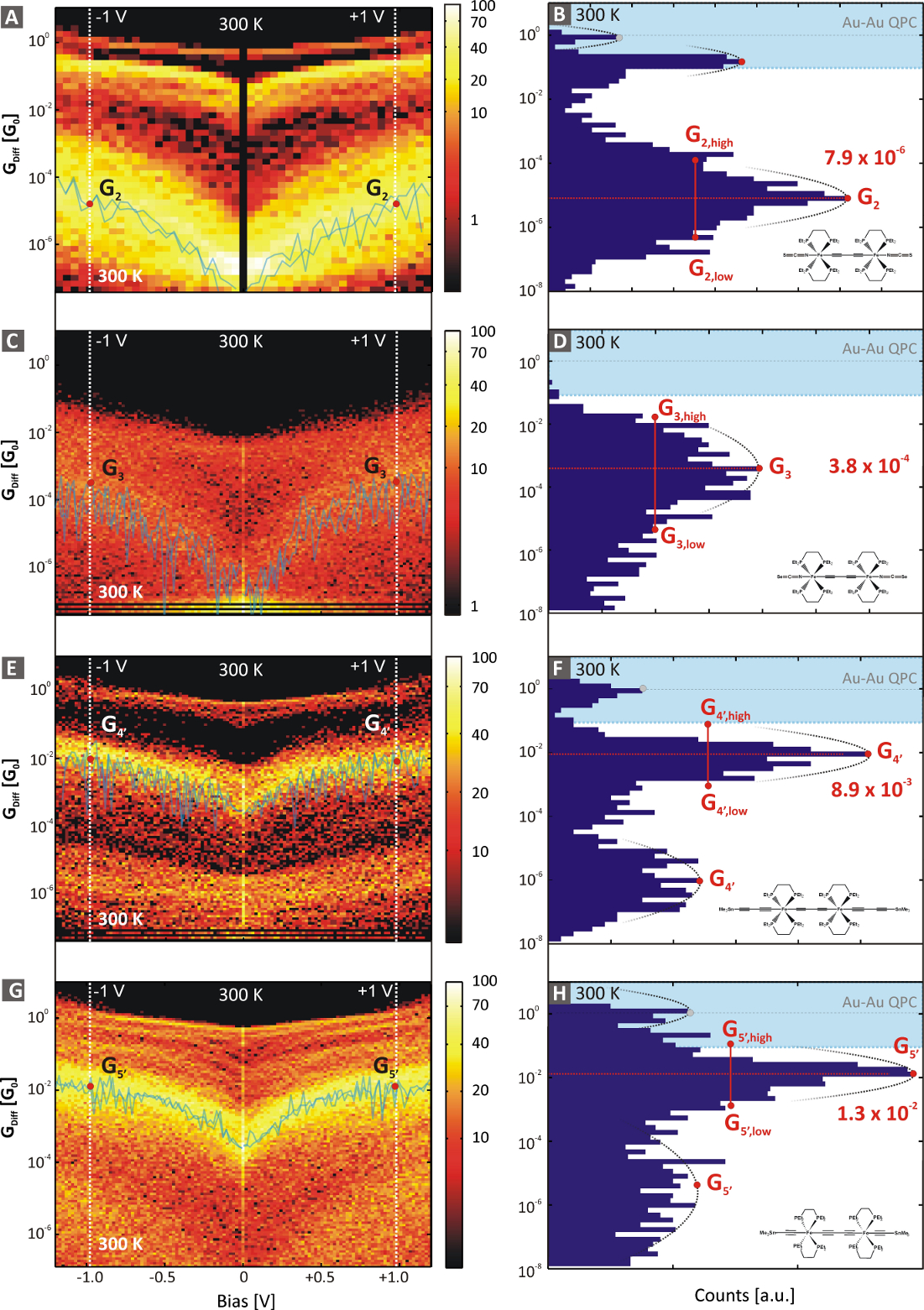}}
    \caption{\label{fig:2} Density plots of the differential conductance, vs.
    voltage, $G_{Diff}$--$V$, characteristics acquired upon repeated opening
    of the junction at 300 K
    for compounds \textbf{2} (A), \textbf{3} (C),
    \textbf{4$'$} (E), and  \textbf{5$'$} (G) upon opening of the junction
    at 300 K.\ Individual $G_{Diff}$--$V$ curves (raw data) are plotted in transparent
    blue
    to display the functional behavior of an individual curve.\
    Corresponding conductance histograms extracted at $\pm$ 1.0 V
    are displayed for \textbf{2} (B),
    \textbf{3} (D), \textbf{4$'$} (F), and \textbf{5$'$} (H).\ The blue area
signals the smallest electrode
    separations that can either lead to a direct Au-Au contact
    (and hence a QPC)
or multi-molecule junctions.\ The maximum conductance accumulation
is labelled in red with a FWHM estimation for the peak width.}
\end{center}
\end{figure*}

\begin{figure*}[htbp]
    \begin{center}
    \resizebox{1.6\columnwidth}{!}{\includegraphics{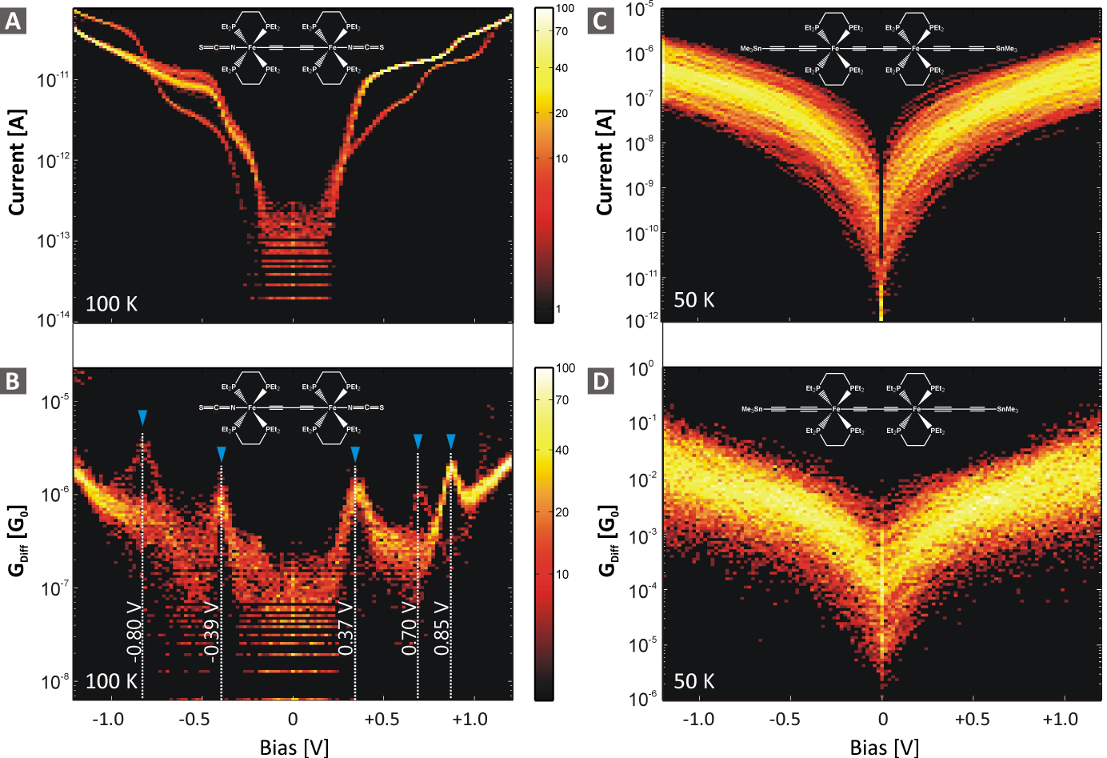}}
    \caption{\label{fig:3} $I$--$V$ and $G_{Diff}$-$V$ characteristics taken at
low temperatures upon repeated opening the junction
    for \textbf{2} in A) and B), and for \textbf{4$'$} in C) and D),
respectively.\ For \textbf{2}, resonant transport
    through molecular orbitals gives rise to conductance peaks at specific
voltages that are symmetric in
    respect to bias.\ In contrast, \textbf{4$'$} reveals exclusively monotonic
curves without
    the appearance of discrete MOs.\ Furthermore, current levels are 3 orders of
magnitude higher for the
    high-bias regime of \textbf{4$'$}, and 4 orders of magnitude higher for the
low-bias regime due to the appearance of a conductance
    gap of approx.\ 0.8 V for \textbf{2}.}
\end{center}
\end{figure*}

\begin{figure*}[htbp]
    \begin{center}
    \resizebox{1.2\columnwidth}{!}{\includegraphics{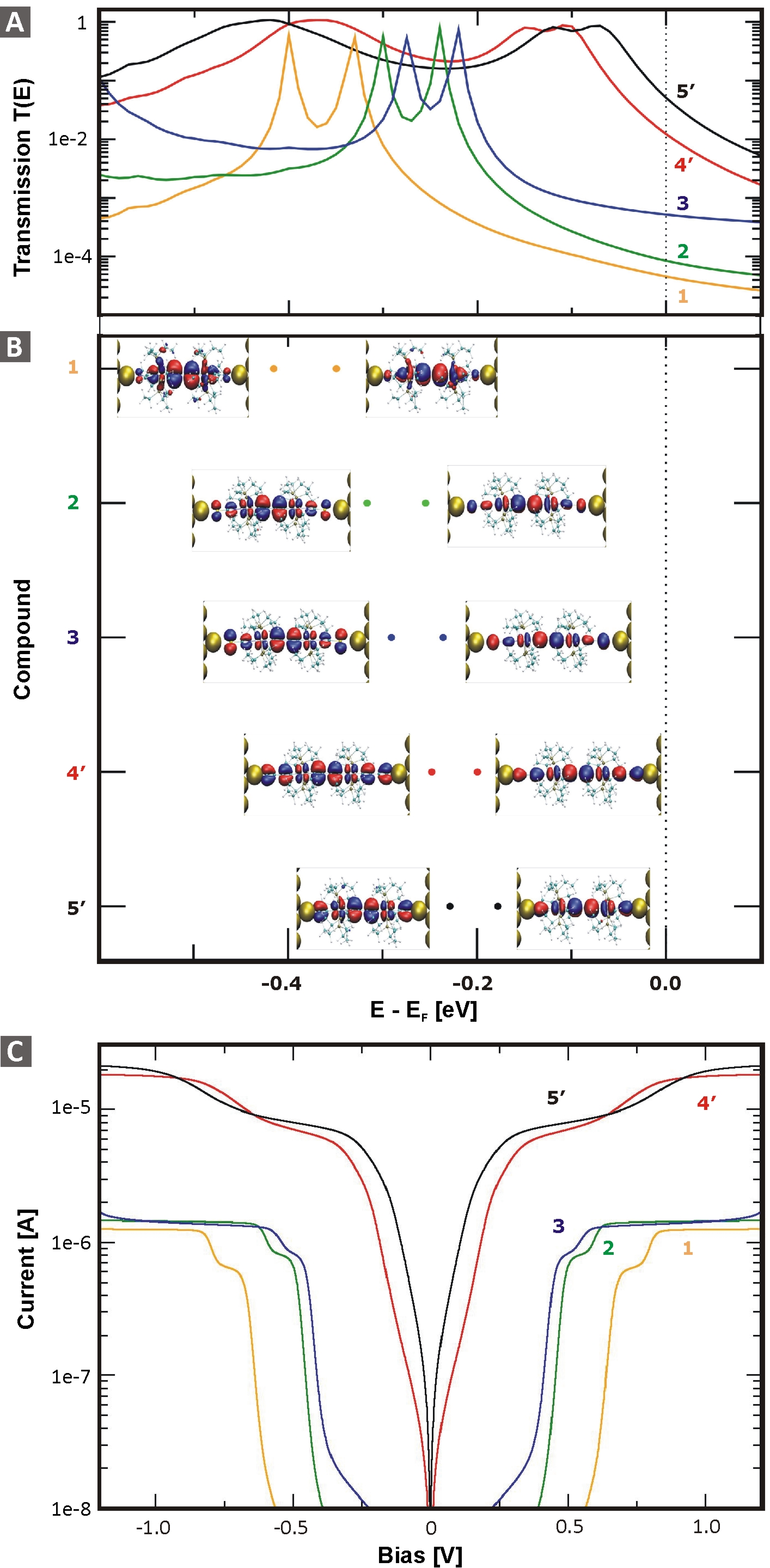}}
    \caption{\label{fig:4} A) Transmission functions for compounds \textbf{1} to
\textbf{5$'$} as calculated
    by DFT (color coding according to B)).\  B) Energetic positions of the
    HOMO and HOMO-1 of compounds \textbf{1} to \textbf{5$'$} represented as dots
with different colors for the different systems with respect to
    the Fermi energy of the electrodes.\ Also given are the
    respective spatial distributions of these HOMO and HOMO-1.\
    The slight shift of the transmission peaks toward the electrode Fermi
    Level results from the hybridization of the MOs with the gold bands,
    which is removed by the subdiagonalization process used to obtain
    the molecular states in the composite system.\ C)
     Calculated $I$--$V$ curves obtained from the transmission
     functions T(E) in a rigid band approximation where the bias
     dependence of T(E) is disregarded.}
\end{center}
\end{figure*}

\begin{figure*}[htbp]
    \begin{center}
    \resizebox{1.6\columnwidth}{!}{\includegraphics{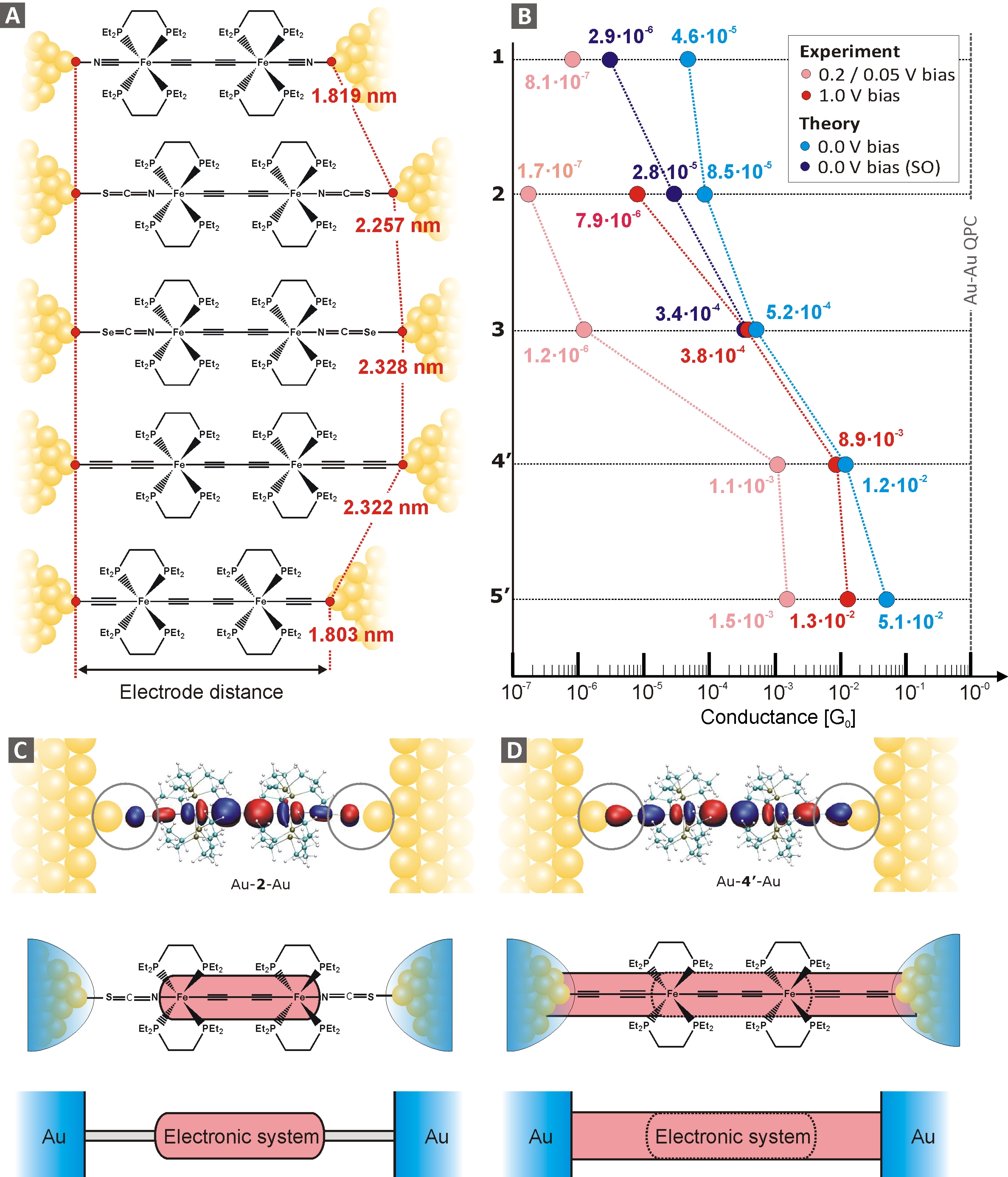}}
    \caption{\label{fig:5} A) Calculated Au--Au distances of the resulting molecular junctions for compounds \textbf{1} to \textbf{5$'$}.\ B)     Comparison of conductances for all compounds determined by experiment (300 K; 200 mV, 1.0 V) and DFT (0 K, zero bias, with and without
    scissor operator (SO) corrections).\
    The experimental data point for \textbf{1} was achieved by low-bias measurements (50 mV).\
    Schematic representation of the Au-\textbf{2}-Au (C) and the Au-\textbf{4$'$}-Au junction (D).\
    The strong hybridization of metal and molecular states in the case of Au-\textbf{4$'$}-Au as
    evidenced by the difference in the HOMO's amplitude on the bonding site as obtained from DFT
    (gray circles),
    leads to the formation of a strong molecule--metal bond and enables to extend the
    delocalized electronic system between the two Fe centers over the molecule--electrode
    interfaces, in contrast to the weakly bonded Au-\textbf{2}-Au system.}
\end{center}
\end{figure*}

\end{document}